\documentclass[%
 reprint,
 amsmath,amssymb,
 aps,
prb,
]{revtex4-2}

\usepackage{amsmath}
\usepackage{todonotes}
\usepackage{graphicx}
\usepackage{dcolumn}
\usepackage{bm}
\usepackage{xcolor}
\usepackage{hyperref}
\usepackage[mathlines]{lineno}
\usepackage[normalem]{ulem}
\usepackage{xr}

\definecolor{modernorange}{HTML}{FF6B35}
\definecolor{darkblue}{HTML}{0066CC}
\definecolor{tealmodern}{HTML}{006D77}
\definecolor{slate}{HTML}{37474F}  
\definecolor{plum}{HTML}{6A4C93}

\newcommand{\OS}{\color{darkblue}}

\begin{document}
\preprint{APS/123-QED}

\title{
Dynamic charge oscillations in a quantum conductor driven by ultrashort voltage pulses}

\author{Lucas Mazzella}
\author{Seddik Ouacel*}
\affiliation{Université Grenoble Alpes, CNRS, Grenoble INP, Institut Néel, Grenoble 38000, France%
}%
\author{Inès Safi}%
 \email{mohamed-seddik.ouacel@neel.cnrs.fr}
\affiliation{%
Laboratoire de Physique des Solides, CNRS UMR 5802-Université Paris-Sud, Université Paris-Saclay, France
}%


\date{\today}
\OS
\begin{abstract}
Ultrashort voltage pulses drive quantum conductors into a nonadiabatic regime where the transmitted charge can oscillate with the injected pulse charge. This effect, to which we refer as dynamic charge oscillations, has been primarily associated with interferometric devices and attributed to path interference. Remarkably, we demonstrate that this constitutes a universal phenomenon, expected to emerge in any quantum conductor that exhibits a sublinear DC current at large bias. Dynamic charge oscillations therefore arise well beyond interferometers and remain perturbatively robust in strongly correlated conductors, including in the presence of arbitrarily strong Coulomb interactions. As a striking example, we predict this effect in a fractional quantum Hall quantum point contact, demonstrating that it can occur in a genuinely noninterferometric system. We further reinterpret the phenomenon in terms of the photoassisted probabilities generated by the voltage pulse.

\end{abstract}

\maketitle

Recent advances in electron quantum optics have enabled the on-demand
generation and coherent control of single-electron excitations,
allowing the transposition of quantum optics paradigms to condensed
matter systems through the manipulation of electrons in ballistic
conductors~\cite{Bocquillon2014, Bauerle2018,Chakraborti_2024}.

Pioneering works in this field include the theoretical prediction of the controlled injection of
electrons using voltage pulses
~\cite{Safi1999}.
In particular, Lorentzian voltage pulses produce single-electron minimal
excitations (so called Levitons)~\cite{Ivanov1997,Keeling2006,Safi2022}
demonstrated experimentally through
shot-noise measurements~\cite{Dubois2013}. In addition, quantum state tomography has provided a full reconstruction of their coherence properties \cite{Jullien2014,Bisognin2019}, while time-resolved techniques have enabled the direct observation of propagating single-electron wave packets and the determination of their speed\cite{Roussely2018,Aluffi2023,Takada2025}.
 
These advances have established precise control over single-electron excitations in ballistic conductors and paved the way for investigations of quantum interference effects and novel dynamical aspects \cite{Gaury2014}. 
Recent experiments have demonstrated interference with Leviton pulses and their coherent manipulation in Mach–Zehnder interferometer (MZI) \cite{Assouline2023}. Related works \cite{Bartolomei2025, Souquet2024} further showed that similar interference effects can be exploited in Fabry–Pérot interferometers for quantum sensing applications. In addition, it has been shown that injecting 30 ps pulses in a MZI enables access to a nonadiabatic regime where internal timescales of the device are probed \cite{Ouacel2025}. These developments have brought the dynamical aspects of quantum transport within experimental reach.
One striking dynamical effect is the oscillation of the transmitted charge as a function of the injected pulse charge when interferometeric systems are driven with ultrashort voltage pulses \cite{Gaury2014,Weston2016, Kloss2025, Saha2025}. 
In what follows, we will refer to this effect as dynamic charge oscillations.
Originally predicted for Fabry–Pérot and Mach–Zehnder interferometers \cite{Gaury2014}, dynamic charge oscillations were later extended to tunnel-coupled wires \cite{Weston2016}.
More recently, a theoretical work showed that similar effects are also
expected in quantum dots, which they still interpret as a Fabry-Perot cavity with large level spacing~\cite{Kloss2025}. 
So far dynamic charge oscillations have not been studied in presence of Coulomb interaction except in  \cite{Prasoon2026} which is restricted to weak interaction.

In this work, we propose a general derivation of dynamic charge oscillations for a generic quantum conductor, extending beyond interferometers and, notably, including strong Coulomb interactions or superconducting correlations. 
We build our approach on the expression of the photo-assisted current valid both for non-interacting systems with arbitrary transmission \cite{Platero2004,Gaury2014_2} and arbitrary strongly correlated circuits in the weak-transmission regime, whose description falls within the Unifying Non-Equilibrium Perturbative (UNEP) approach \cite{Safi2019}.
We show that dynamic charge oscillations are expected  in any quantum conductor with a sublinear DC characteristic.
Thus, we propose a complementary interpretation that originates from oscillations of the photo-assisted probabilities associated with the AC excitation, rather than the traditional interferometry picture. To further support our argument, we study a non-interferometric system  subject to a Lorenztian pulse with a finite width, a quantum point contact in the fractional quantum Hall (FQH) regime. It leads to clear dynamic charge oscillations in the short pulse limit, supporting our claim that the sublinearity of the DC characteristic is sufficient to observe dynamic charge oscillations.

\paragraph{\label{sec:sec2}Quantum conductor in the short-pulse limit}

We start from a minimal description of a generic quantum conductor, characterized by its DC current–voltage relation $I_{\mathrm{dc}}(\omega)$ where $\omega = \frac{e V}{\hbar}$ is a frequency associated with the DC voltage $V$. 
When driven by a time-dependent voltage pulse $V(t)$, the Fermi sea acquires an additional phase $\varphi(t) = \frac{e}{\hbar}\int_{-\infty}^t dt'V(t')$ that corresponds to the injection of a total charge
\begin{equation}
\label{eq:charge}
  q = \int_{-\infty}^{+\infty} dt  \frac{eV(t)}{h}=\frac{\varphi(t=+\infty)}{2\pi}.  
\end{equation}

\noindent For simplicity, we have written the Josephson-type relations obeyed by $\omega$ and $\varphi(t)$ for a single channel of independent electrons with charge $e$, noting that the formalism readily generalizes to fractional charges, Cooper pairs or multichannel systems \cite{Safi2019}.

In the following, we focus on the ultrashort-pulse limit, where the pulse duration is shorter than any internal timescale of the conductor. In that regime, the drive can be approximated as $V(t)=V\,\delta(t)$, leading to a phase $\varphi(t)=2\pi q\,\theta(t)$, where $\delta(t)$ is the Dirac delta function and $\theta(t)$ is the Heaviside step function.
The excitation generated by the pulse is encoded in the Fourier transform of the phase factor, which defines the photoassisted amplitudes
\begin{equation}
\label{eq:p_omega}
  p(\omega)=\int_{-\infty}^{+\infty} e^{-i\varphi(t)}e^{i\omega t}\,dt ,  
\end{equation}
\noindent whose squared modulus $|p(\omega)|^2$ (or, for periodic drives, the discrete weights $|p_l|^2$with integer $l$) gives the probability to absorb or emit a photon at frequency $\omega$   \cite{Moskalets2002, Dubois2013_2}.
We study the transmitted charge $\bar{n}$, which can be expressed through the photo-assisted relation
\begin{equation}
\label{eq1:current_photo_nonperiodic}
e\bar{n} = \int_{-\infty}^{+\infty} \frac{d\omega}{2\pi}\, |p(\omega)|^2\, I_{\mathrm{dc}}(\omega).
\end{equation}
\noindent For non-interacting conductors, Eq.\,\eqref{eq1:current_photo_nonperiodic} is exact ~\cite{Gaury2014_2} and has been used in all previous derivations of dynamic charge oscillations~\cite{Rossignol2018,Weston2016,Kloss2025}. 
Remarkably, it has been shown to hold in full generality within the UNEP framework for strongly correlated circuits (see End Matter \ref{sec:UNEPT}).
 This approach extends the side-band transmission picture for single electron states under a periodic voltage drive \cite{Platero2004} to non-periodic drives with which many-body correlated states exchange an energy $\hbar\omega$   \cite{Safi2019} with  probabilities $ |p(\omega)|^2 $. These probabilities  weight the DC characteristics 
$ I_{\mathrm{dc}}(\omega)$ that carry alone the signature of strong correlations.
Since we study ultrashort pulses, Eq.~\eqref{eq1:current_photo_nonperiodic} diverges if $I_{\mathrm{dc}}(\omega)$ is linear. This divergence is a manifestation of the orthogonality catastrophe \cite{Lee1993}. To avoid this issue, we assume a sublinear behavior of the DC characteristics at large bias, namely
\begin{equation}\label{eq:sublinearity}
\lim_{\omega\to\pm\infty}\frac{I_{\mathrm{dc}}(\omega)}{\omega}=0.
\end{equation}


Eq.\,\eqref{eq1:current_photo_nonperiodic} serves as our starting point to evaluate the transmitted charge in the short-pulse limit. For that, we introduce a standard regularization:

\begin{equation}
\begin{aligned}
p_\epsilon(\omega)=\int_{-\infty}^{+\infty} e^{-i\varphi(t)}e^{i\omega t}e^{-\epsilon|t|}\,dt 
\\
e\bar{n}_\epsilon=\int_{-\infty}^{+\infty} \frac{d\omega}{2\pi}\,|p_\epsilon(\omega)|^2\,I_{\mathrm{dc}}(\omega)
\end{aligned}
\end{equation}
such that $\lim_{\epsilon\to0}\bar{n}_\epsilon=\bar{n}$. With $\varphi(t)=2\pi q\,\theta(t)$ one finds
\begin{equation}
\label{eq2:P_w_regularized}
|p_\epsilon(\omega)|^2=\frac{2}{\omega^2+\epsilon^2}
-\frac{e^{2i\pi q}}{(\omega-i\epsilon)^2}
-\frac{e^{-2i\pi q}}{(\omega+i\epsilon)^2}.
\end{equation}
\noindent Substituting into Eq.\,\eqref{eq1:current_photo_nonperiodic} gives three terms. For the last two, an integration by parts yields
\[
\int_{-\infty}^{+\infty} d\omega\,\frac{I_{\mathrm{dc}}(\omega)}{(\omega\pm i\epsilon)^2}
=\int_{-\infty}^{+\infty} d\omega\,\frac{G_{\mathrm{dc}}(\omega)}{\omega\pm i\epsilon},
\]
where $G_{\mathrm{dc}}(\omega)=\frac{dI_{\mathrm{dc}}}{d\omega}(\omega)$ is the rescaled conductance that matches the transmission probabilities up to a factor $2\pi$ in the non-interacting case. 
The boundary term
$[I_{\mathrm{dc}}(\omega)/\omega]_{-\infty}^{+\infty}$ vanishes due to the sublinearity hypothesis formulated in Eq.\eqref{eq:sublinearity}.
Taking the limit $\epsilon\to0$ and using the Sokhotski–Plemelj formula then gives:
\begin{equation}
\label{eq3:I_ph_final}
e\bar{n} = \frac{1}{\pi}\,\mathrm{P.V.}\!\int_{-\infty}^{+\infty} d\omega\,\frac{G_{\mathrm{dc}}(\omega)}{\omega}\,\big[1-\cos(2\pi q)\big]
\;+\; G_{\mathrm{dc}}(0)\,\sin(2\pi q),
\end{equation}
 \noindent where P.V. denotes the principal value. Eq.\,\eqref{eq3:I_ph_final} relies on only two assumptions: 
(i) the photo-assisted form of Eq.\,\eqref{eq1:current_photo_nonperiodic}, whose domain of validity has already been discussed and 
(ii) the sublinearity of the DC characteristic. Note that, an additional constant contribution $G_0$ to $G_{\mathrm{dc}}(\omega)$ can be treated separately and would simply add a trivial term $G_0 q$ to Eq.\,\eqref{eq3:I_ph_final}.

Our derivation is independent on the microscopic details of the system. This unifies the predictions of dynamic charge oscillations reported in previous works in four distinct non-interacting systems~\cite{Rossignol2018,Weston2016,Kloss2025}: Fabry-P\'erot interferometers, MZIs, tunnel-coupled wires, and quantum dots. A detailed verification is provided in Supplementary material, demonstrating that our expression in Eq.\,\eqref{eq3:I_ph_final} reproduces all these results in the ultrashort pulse limit. 
In those works, dynamic charge oscillations are interpreted as an interference phenomenon between different propagating paths.
Strikingly, our derivation extends to non-interferometric devices, showing that this is a universal phenomenon in quantum conductors driven with ultrashort voltage pulses.

In what follows, we illustrate and verify this generalization by analyzing a quantum point contact (QPC) in the fractional quantum Hall (FQH) regime driven by a Lorentzian pulse of finite width.

\begin{figure}[h!]
\begin{center}
\includegraphics[scale=1]{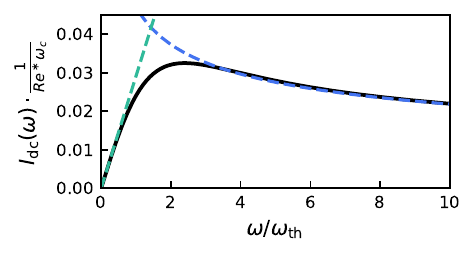}
\caption{DC backscattering current at the  QPC with $R$ = 0.01, $\nu=\delta= 1/3$  and ${\omega_{\mathrm{th}}} = 0.01\cdot\omega_c$. 
The green dotted line represents the equilibrium regime $\omega \ll \omega_{\mathrm{th}}$. 
The blue dotted line represents the non-equilibrium regime $ \omega\gg \omega_{\mathrm{th}}$.}
\label{fig_2}
\end{center}
\end{figure}

\paragraph{\label{sec:sec4} Quantum point contact in the fractional quantum Hall regime}
\begin{figure}[h!]
\includegraphics[scale=1]{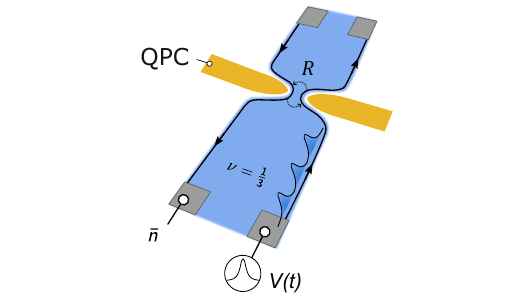}
\caption{Schematic of a QPC in the FQH edges. Electrons are injected by a periodic train of voltage pulses $V(t)$ applied to an Ohmic contact, inducing the backscattered charge $\bar{n}$.}
\label{fig_1}
\end{figure}
In this section, we consider Laughlin FQH edges at simple filling factor $\nu$ described by a Tomonaga-Luttinger Liquid with a weakly backscattering QPC (see  Fig.\,\ref{fig_1}\textbf{a}). The DC backscattering current can be written as \cite{Wen1991}
\begin{align}
\label{eq:i_qpc}
{I_{\mathrm{dc}}(\omega)=
\,R\,e^\ast\,
\frac{\omega_{\mathrm{th}}}{\pi\Gamma(2\delta)}\,
\left(\frac{\omega_{c}}{4\pi^2\omega_{\mathrm{th}}}\right)^{2(1-\delta)}\,}
\\[6pt] \notag
\times \left|\Gamma\!\left(\delta + i\, \frac{\omega}{2\pi\omega_{\mathrm{th}}} \right)\right|^{2}
\sinh \bigg( \frac{\omega}{2\omega_{\mathrm{th}}} \bigg ),
\end{align}
\noindent where $\omega\!=\!e^\ast V/\hbar$ with $e^\ast$ the quasiparticle charge, 
$\delta$ the scaling dimension, $\omega_{\mathrm{th}}\!=\!k_B T/\hbar$ the thermal frequency, $R$ the QPC reflection  coefficient, and $\omega_{\mathrm c}$ a high–energy cutoff.
To be specific, we choose $\nu=\delta=e^\ast/e=1/3$ and $R=0.01$. We also fix a thermal frequency ${\omega_{\mathrm{th}}} = 0.01\,\omega_c$ to remain within the validity domain of the weak–backscattering description~\cite{Safi2025} (see End Matter\,\ref{B1: limit TLL}).
The resulting DC backscattering current is shown in Fig.~\ref{fig_2}. 
The thermal frequency $\omega_{\mathrm{th}}$ separates two distinct regimes. 
For $\omega \ll \omega_{\mathrm{th}}$, the energy injected by the DC drive is smaller than the thermal energy. Quasiparticle excitations with energies of order $k_B T$ are already thermally populated at the 
QPC, so that backscattering processes are effectively thermally activated. The applied voltage merely biases 
these pre-existing fluctuations, leading to a  linear voltage bias characteristic, as shown by the green dashed line. 
In contrast, for $\omega \gg \omega_{\mathrm{th}}$, quasiparticles are emitted at energies fixed by the DC drive. Thermal fluctuations are irrelevant at this energy scale.  Therefore, the QPC probes the intrinsic quantum correlations of the fractional edge. The resulting backscattering current 
exhibits a non-linear power-law dependence on the bias, characteristic of chiral Luttinger-liquid behavior (see the blue dashed line).
It clearly satisfies the sublinearity condition in Eq.\eqref{eq:sublinearity}, which, as noted above, is the key condition for observing dynamic charge oscillations in the ultrashort pulse limit.
Having established the DC response and its sublinearity, we now consider a periodic train of Lorentzian voltage pulses of width $\tau$ and period $T_0$,
\begin{equation}
\label{eq:lorentzian}
 e V(t)=\sum_{n=-\infty}^{+\infty} \frac{2\,q\,\hbar\,\tau}{\tau^2+(t-nT_0)^2},
\end{equation}
{which injects a charge $qe$ per pulse. The single–pulse limit is recovered for $T_0\to\infty$ at fixed $\tau$. We write Eq.\,\eqref{eq1:current_photo_nonperiodic} in the periodic case} \cite{Platero2004,Safi2019},
\begin{equation}
\label{eq:floquet_qpc}
e\bar{n}
= \frac{2\pi}{\Omega_0} \sum_{l=-\infty}^{+\infty} |p_l|^2 \, I_{\mathrm{dc}}\left [(l+q\,)\Omega_0\right], \qquad
\Omega_0=\frac{2\pi}{T_0},
\end{equation}
where
\[
{p_l=\frac{\Omega_0}{2\pi}\int_{-T_0/2}^{T_0/2} e^{-i\varphi(t)} e^{i l \Omega_0 t}\,dt}
.\]
The shift $l\!+\!q$ reflects the fact that we remove the DC component of $V(t)$ when defining the phase $\varphi(t)$.
Eq.\,\eqref{eq:floquet_qpc} is valid provided that we stay in the perturbative regime and within the validity domain of Eq.\,\eqref{eq:i_qpc} as discussed in {End Matter} \ref{B1: limit TLL} and \ref{B2: limit unept}. 

\begin{figure}[h!]
\begin{center}
\includegraphics[scale=1]{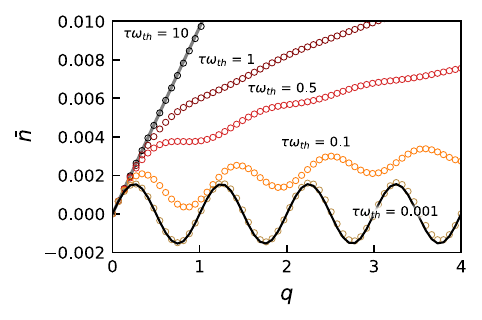}
\caption{Backscattered charge $\bar{n}$ (given by Eq.\,\eqref{eq:floquet_qpc}) as a function of the injected charge $q$ for Lorentzian pulses of various widths $\tau$ at $R$ = 0.01, $\nu$= 1/3, $\delta$ = 1/3, ${\omega_{\mathrm{th}}} = 0.01\,\omega_c$ and $\Omega_{0} = 10^{-3}\,\omega_c$. Solid gray line: adiabatic limit. Solid black line: ultrashort pulse limit given by Eq.\,\eqref{eq3:I_ph_final}.
}
\label{fig_3}
\end{center}
\end{figure}

Eq.\,\eqref{eq:floquet_qpc} is evaluated numerically, and the results are displayed in Fig.~\ref{fig_3} for different values of pulse width $\tau\omega_{\mathrm{th}}$. 
We choose the repetition frequency as $\Omega_0 = 0.1\,\omega_{\text{th}}$, such that 
$\Omega_0 \ll \omega_{\mathrm{th}}$ and $\Omega_0 \ll 1/\tau$. 
These conditions ensure that pulses are well separated in time and that the system fully relaxes 
between successive pulses. In this limit, the periodic drive effectively reduces to the single-pulse regime, allowing direct comparison of the numerical results with Eq.~\eqref{eq3:I_ph_final}.
For $\tau\omega_{\mathrm{th}} \gg 1$, the backscattered charge $\bar{n}$ follows the adiabatic limit (gray solid line in Fig.~\ref{fig_3}), where the voltage varies slowly compared to all intrinsic timescales of the conductor (see End Matter\,\ref{B3:adiabatic limit}).
When $\tau\omega_{\mathrm{th}} \sim 1$, dynamic charge oscillations of $\bar{n}$ as a function of the injected charge $q$ appear and become more pronounced as the pulse width $\tau$ is reduced. 
In the ultrashort-pulse limit $\tau\omega_{\mathrm{th}}\ll 1$, the results are captured by the analytical expression of Eq.\,\eqref{eq3:I_ph_final} (black solid line in Fig.~\ref{fig_3}). Notice that the analysis of the photo-assisted current in \cite{Martin2017}  exhibits early signatures of oscillatory behavior at intermediate $\tau\omega_{\mathrm{th}}$ (e.g., $\tau\omega_{\mathrm{th}}\sim 0.1$), although without explicitly linking this effect to dynamic charge oscillations. 
 In the QPC, the crossover displayed in Fig.~\ref{fig_3} as a function of $\omega_{\mathrm{th}}$ reproduces the behavior reported in interferometric devices: there, $1/\omega_{\mathrm{th}}$ effectively replaces the  time of flight imbalance between paths~\cite{Rossignol2018,Weston2016}, or the finite lifetime of a quantum-dot level~\cite{Kloss2025}.
 This behavior can be interpreted in the time domain by introducing the thermal coherence time
$\tau_{\mathrm{th}} = 1/\omega_{\mathrm{th}}$, beyond which  quasiparticle correlations are
suppressed by thermal fluctuations. For long pulses, $\tau \gg \tau_{\mathrm{th}}$, thermal fluctuations
destroy coherence while the pulse propagates through the QPC, thereby smearing out the
oscillations. In contrast, for short pulses, $\tau \ll \tau_{\mathrm{th}}$, coherence is maintained throughout
the passage of the pulse, allowing dynamic charge oscillations to emerge. Thereby, we demonstrate that dynamic charge oscillations emerge in a non-interferometric device, matching the ultrashort-pulse prediction of Eq.~\eqref{eq3:I_ph_final}. As a concrete estimation, at an electronic temperature of $T=30$\,mK, dynamic charge oscillations should be measurable for pulse width shorter than $\tau_{th}=250$\,ps which is experimentally accessible.

\paragraph{\label{sec:sec5}Interpretation of dynamic charge oscillations}
The standard interpretation of dynamic charge oscillations relies on interference between different propagating paths \cite{Gaury2014, Saha2025}. 
However, since we predict dynamic charge oscillations in non-interferometric systems as well, a more general interpretation applicable to both cases is required.
A closer inspection of the expression in Eq.\,\eqref{eq2:P_w_regularized} reveals that the oscillating term originates directly from the phase factor.
This feature is inherent to the excitation generated by the voltage pulse and does not depend on the specific quantum system under consideration.
Our interpretation is therefore that the origin of dynamic charge oscillations lies in the oscillatory behavior of photo-assisted probabilities, which we are able to probe thanks to the sublinearity of the device. 
Interference is one possible mechanism leading to sublinear DC characteristic, but other 
energy-dependent processes can play the same role, such as tunneling in the case of 
a QPC in the FQH regime.
This explains why dynamic charge oscillations manifest in different  systems.

To illustrate our interpretation,  we write the expression of the photo-assisted probabilities $|p_l|^2$ in the short-pulse limit for a periodic train of pulses \cite{Safi2019} (see End Matter \ref{A4:Pl_short} for the derivation):
\begin{equation}
\label{eq8:pn_short}
    |p_l|^2 = \frac{\text{sin}^2(\pi q)}{\pi^2( l+ q)^2}
\end{equation}
\noindent First, the oscillations with the injected charge $q$ appear in the expression of $|p_l|^2$ as it was the case for the non-periodic case in Eq.\,\eqref{eq2:P_w_regularized}.
In the non-interacting case, $|p_l|^2$ represents the probability to exchange $|l|$  photons with the pure AC source which shifts a single electron energy state from  $q\Omega_0$ to the Floquet state $(l+q)\Omega_0$.
This simple picture can be extended to interacting systems by considering many-body correlated states instead of single electrons states \cite{Safi2019}.
We also consider a symmetric DC characteristic $I(\omega)=-I(-\omega)$ such that Eq.\,\eqref{eq3:I_ph_final} reduces to 
\[
\bar{n}=G_{\mathrm{dc}}(0)\,\sin(2\pi q)
\]
When $q$ is an integer, Eq.\,\eqref{eq8:pn_short} becomes $|p_l|^2 = \delta_{l,-q}$.  
One is left with a single term in the sum in Eq.\,\eqref{eq:floquet_qpc} with $|p_{-q}|^2 = 1$ leading to no transferred charge $\bar{n}$ because the Floquet state has energy $(l+q)\Omega_0 = 0$.

When $q = m + \frac{1}{2}$, Eq.\,\eqref{eq8:pn_short} gives $|p_l|^2 = \frac{1}{\pi^2 \left(l + m + \frac{1}{2}\right)^2}$.  
In this case, $|p_{l-m}|^2 = |p_{-l-m-1}|^2$ which means equal probabilities to populate the Floquet states with energies $(l+1/2)\Omega_0$ and $-(l+1/2)\Omega_0$. 
The contributions of those two states to the transferred charge $\bar{n}$ in Eq.\,\eqref{eq:floquet_qpc} cancel each other because the DC current is odd such that $\bar{n} = 0$.
The above considerations show that dynamic charge oscillations can be understood by considering the
balance between the different Floquet states. It reinforces our claim that the
origin of this phenomenon lies in the AC drive itself.

To conclude, we have demonstrated the universality of oscillations of the transmitted  charge through quantum conductors (so called dynamic charge oscillations)  as a function of the charge injected by ultrashort voltage pulses. Our analysis unifies, on one hand, non-interacting systems with arbitrary energy-dependent transmission amplitudes, and, on the other hand, strongly correlated systems or circuits described by the  Unified Non Equilibrium Perturbative (UNEP) theory.  Remarkably, these oscillations extend beyond interferometers: they emerge as soon as the DC current-voltage characteristic is sublinear at large voltage, irrespective of the microscopic details of the system.
\\
We have verified our results through a numerical analysis of a Lorentzian pulse with finite width applied to a quantum point contact (QPC) in {fractional quantum Hall} (FQH) edges. By assuming the validity of the Tomonaga-Luttinger liquid  model in the weak backscattering regime for which the sublinearity condition is satisfied, we capture the crossover from the adiabatic to the short-pulse regime and demonstrate the emergence of dynamic charge oscillations in the short-pulse limit. Notice that our approach extends to hierarchical FQH states as long as backscattering of a single quasiparticle species dominates. While the DC current keeps the same form \cite{Safi2025_2} in the presence of edge interactions, we expect, however, that the latter induce fractionalization of the injected pulse \cite{Safi1999,Levkivskyi2008,Kamata2014} whose spreading needs therefore to be reduced.

To date, these charge oscillations have not been observed experimentally. Our results provide a practical roadmap to identify platforms in which such dynamic charge oscillations could be detected on experimentally accessible timescales. Moreover, our framework can be extended to an initial non-equilibrium distribution due, for instance, to temperature gradients or three-terminal geometries and to hybrid systems with superconducting correlations (e.g., proximitized junctions), where Andreev processes and the BCS density of states naturally generate nonlinear DC responses. 

Interestingly, our photo-assisted approach could be extended to the time-resolved current, thereby revealing dynamical charge oscillations directly in the time domain. From this viewpoint, these oscillations can be rooted into interference effects analogous to time-domain braiding processes between anyons, as interpreted in "anyon-collider" \cite{Han2016,Lee2020} or Hong-Ou-Mandel-type setups \cite{Ruelle2025,Jonkheere2023,*Matteo2025,*Latyshev2025}.
\begin{acknowledgments}
The authors acknowledge fruitful discussions with Pascal Degiovanni, Gwendal Fève and Aleksander Latyshev during the preparation of this manuscript. They thank Christopher B\"auerle, Thomas Vasselon, Matteo Aluffi and Corentin Deprez for carefully reading the manuscript and for valuable remarks. I.S. acknowledges the ANR grant "QuSig4QuSense" (ANR-21-CE47-0012). S.O. acknowledges funding from the European Innovation Council and SMEs Executive Agency under Grant Agreement 101185712, project ELEQUANT. L.M. acknowledges the program QuanTEdu-France No. ANR-22-CMAS-0001 (France 2030).

Views and opinions expressed are those of the author(s) only and do not necessarily reflect those of the European Union or the granting authority. Neither the European Union nor the granting authority can be held responsible for them.
\end{acknowledgments}
\bibliography{main_bib}
\appendix
\newpage
\section*{End matter}

\subsection{Unifying Non Equilibrium Perturbative theory}
\label{sec:UNEPT}

Here, we briefly recall the Unifying Non Equilibrium Perturbative framework. For a more detailed derivation, see\cite{Safi2019,Safi2022,Safi2025}.

We start from the following Hamiltonian

\begin{eqnarray}
    \label{Hamiltonian} 
	H(t)&=&H_0 +H_{A}(t),\nonumber\\
	H_{A}(t)&=&e^{-i\omega t} p(t){A}+p^*(t)e^{i\omega t}{A}^{\dagger} 
\end{eqnarray}
where $H_0$ is the time-independent Hamiltonian which can contain arbitrary strong interactions and $H_{A}$ is a time-dependent perturbation to $H_0$ due to the AC drive.  
$\omega$ is a free frequency parameter that encodes the DC drives while AC drives are encoded in the complex function $p(t)=e^{-i\varphi(t)}$ where $\varphi(t) = \frac{e^\ast}{\hbar}\int_{-\infty}^t dt'V(t')$ is the phase associated with the voltage pulse $V(t)$ and $e^\ast$ is the charge which can be fractional. 
The weak operator $A$ obeys the condition $\langle A(t) A(0) \rangle = 0$, where $A(t) = e^{\frac{i}{\hbar}H_0 t} A e^{-\frac{i}{\hbar}H_0 t}$  is taken in the interaction picture where average is defined with respect to a stationary initial distribution.
For example, in the case of FQH edge states, $H_0$ describes the fractional edge states and $A$ corresponds to quasi-particle backscattering operator at the QPC, $A=\zeta \Psi_u(0)^{\dagger}\Psi_d(0)$. Here $\Psi_u$ and $\Psi_d$ are the upper and lower edge quasiparticle fields evaluated at the position $x=0$ of the QPC and $\zeta$ is the backscattering amplitude.

We focus on the renormalized "current operator" associated with the perturbing term:
 $$\hat{I}(t) =\frac{1}{\hbar}\frac{\delta H_{{A}}(t)}{\delta{\varphi}(t)},$$
 which, in view of Eq.(\ref{Hamiltonian}), yields:
\begin{equation}
	\label{eq:current}
	i\hbar\hat{I}(t)\! =\!
		e^{-i\omega t}p(t)\;{A}- e^{i\omega t}p^*(t)\; 
{A}^{\dagger}\,.
\end{equation}  
Average current is given by $I(\omega;t)=\langle \hat{I}_H(t)\rangle
$, where $ \hat{I}_H(t)$ is taken in the Heisenberg representation. 

We can use the generalized linear response theory formulated with respect to nonequilibrium states:
\begin{eqnarray}\label{eq:average_ph}
\langle \hat{I}_H(t) \rangle&=&\frac{-i}{\hbar} \int^{t}_{-\infty} \!\!\!dt^{\prime} \langle [\hat{I}(t), \hat{H}_{A}(t^{\prime})] \rangle\nonumber\\&=&\frac{-1}{\hbar^{2}} \int^{t}_{-\infty} \!\!\!dt^{\prime} e^{i\omega(t^{\prime}-t)}p(t)p^*(t') \langle [{A}(t),{A}^{\dagger}(t^{\prime})] \rangle\!+\!h.c.\nonumber\\&=&p(t)\int^{0}_{-\infty}\!\!d\tau e^{i\omega\tau}p^*(t+\tau) X_{-}(\tau)\!+\!h.c.
\end{eqnarray}
where we have introduced the  commutator that is determined by the stationary dynamics of ${H}_0$: 
\begin{eqnarray}
\label{Xup_Xdown}
X_{-}(\tau)&=&\langle [{A}^{\dagger}(\tau),{A}(0)] \rangle
\end{eqnarray} 
Let us first take  $p(t)=1$ in Eq.\eqref{eq:average_ph}. 
On the one hand, $\langle \hat{I}_H(t) \rangle$ does not depend anymore on time $t$ and reduces to the stationary DC current ${I}_{dc}(\omega)$.
On the other hand, the right-hand term reduces to the Fourier transform $X_{-}(\omega)$ of $X_{-}(\tau)$.
This leads to the following identity
\begin{equation}\label{eq:Idc_omegadc}
 {I}_{dc}(\omega)= X_{-}(\omega).
 \end{equation}
\\ 
Second, we rewrite Eq.\eqref{eq:average_ph} in the presence of $p(t)$ using Eq.\eqref{eq:Idc_omegadc} and decomposing $X_{-}(\tau)$ into its Fourier components, yielding 

\begin{eqnarray}\label{eq:average_ph_2}
\langle \hat{I}_H(t) \rangle\!&\!=&p(t)\!\!\int^{+\infty}_{-\infty}\!\!\frac{d\omega'}{2\pi}{I}_{dc}(\omega')\!\!\!\int^{0}_{-\infty}\!d\tau  e^{i(\omega-\omega')\tau}p^*(t+\tau) \!+\!h.\!c.\nonumber\\&&
\end{eqnarray}

We finally take the average on time $t$. For that, we use the identity
 $$\int dt \;p(t)p^*(t+\tau)=\int \frac{d\Omega}{2\pi}\; |p(\Omega)|^2 e^{i\Omega\tau}$$ where $p(\Omega)$ is the Fourier transform of $p(t)$. Therefore, the average over $\tau$ in Eq.\eqref{eq:average_ph} fixes $\Omega=\omega'-\omega$, thus yielding
\begin{eqnarray}\label{eq:Iph_Xph}
 {I}_{ph}(\omega)&=&\frac{1}{T_0}\int^{+\infty}_{-\infty}\!\! \!\frac{d\omega'}{2\pi} |p(\omega'-\omega)|^2{I}_{dc}(\omega')
 \end{eqnarray}
where $T_0$ is the measurement time.
Using that $I_{ph}=\frac{e\bar{n}}{T_0}$ and taking $\omega =0$ (no additional DC drive), we obtain Eq.\,\eqref{eq1:current_photo_nonperiodic}.

\subsection{Validity of the weak backscattering regime in the fractional quantum Hall (FQH) effect}
\subsubsection{DC regime}
\label{B1: limit TLL}

In order to stay in the weak backscattering regime (Eq.\,\eqref{eq:i_qpc}), the DC drive must verify that the DC backscattering conductance is small with respect to the quantized conductance 
\[
    \frac{e}{\hbar}\, |G_{\mathrm{dc}}(\omega)| \ll \nu\,\frac{e^{2}}{h}.
\]
For $\nu=\delta=e^\ast/e=1/3$ and a reflection coefficient $R=0.01$, this condition is safely ensured with a finite thermal cutoff satisfying $\omega_{\mathrm{th}} > 8\times 10^{-3}\,\omega_c $\cite{Safi2025}. For this reason, throughout this work, we fix the thermal frequency to the conservative value $ \omega_{\mathrm{th}} = 0.01\,\omega_c .$
Another limitation is that the TLL expression of the DC current applies only for an argument below the high-energy cutoff 
$\omega_{c}$. 
Consequently, when evaluating Eq.\,\eqref{eq:floquet_qpc}, we explicitly verify that the sum has converged for all contributions with $\omega < 0.1\,\omega_{c}$. 
\subsubsection{AC regime}
\label{B2: limit unept}
The UNEP framework extends Eq.\,\eqref{eq1:current_photo_nonperiodic} and \eqref{eq:floquet_qpc} to interacting systems. The expressions for $\omega$ and $q$ in terms of external voltages might be more complicated for an initial non-equilibrium distribution, for instance due to temperature gradients or three-terminal geometries, but they are related universally to the effective phase felt by the scattering region.
The conditions for the validity of the UNEP applied to a TLL model are discussed in 
Ref.~\cite{Safi2025}. 
Let us recall them here briefly using our notations.

We define the photo-current as
\[
    I_{\mathrm{ph}} = \frac{e\,\bar{n}}{T_{0}},
\]
and the corresponding photo-conductance as
\[
    G_{\mathrm{ph}} = \frac{\mathrm{d} I_{\mathrm{ph}}}{\mathrm{d} V_{0}},
    \qquad 
    V_{0} = q\frac{\hbar \Omega_{0}}{e^{\ast}}.
\]
The perturbative treatment remains justified as long as
\[
    |G_{\mathrm{ph}}| \ll \nu\,\frac{e^{2}}{h}.
\]

Rewriting this condition in terms of charges yields
\[
|\frac{d\bar{n}}{dq}| \ll 1
\]

 For all the injected charges $q$ and pulse widths $\tau$ we considered in this work, $|\frac{d\bar{n}}{dq}| < 0.03$,  which satisfies this condition.

\subsubsection{Adiabatic limit}
\label{B3:adiabatic limit}
We consider the limit of a long pulse ($\Omega_0\ll 1/\tau\ll \omega_{\mathrm{th}}$), such that the current at time $t$ is simply $I_{\mathrm{dc}}[V(t)]$ (for simplicity, we use the voltage $V$ as a variable for the current instead of $\omega$ as in the rest of this work).

\begin{align*}
e\bar{n}&= \int_0^{T_0} dt \ I_{\mathrm{dc}}[V(t)] \\
\end{align*}
For small $q$, we can verify that $e^\ast V(t)/\hbar \ll \omega_{\mathrm{th}}$ (see Eq.\,\eqref{eq:lorentzian}) such that $I_{\mathrm{dc}}[V(t)]$ becomes linear. 
We get 
\begin{align*}
e\bar{n}&=qe^\ast R\frac{|\Gamma(\delta)|^2}{\Gamma(2\delta)}\bigg( \frac{\omega_c}{4\pi^2\omega_{\mathrm{th}}}\bigg)^{2(1-\delta)} 
\end{align*}

\subsection{Expressions for $|p_l|^2$ in the short-pulse limit \label{A4:Pl_short}}
In this section, we derive the expression of $|p_l|^2$ for a periodic train of pulses in the short-pulse limit given in Eq.\,\eqref{eq8:pn_short}.
We start with the expression of $p_l^{\eta}$ for a periodic train of Lorentzian pulses \cite{Dubois2013}.
\[
p_l^{\eta} = \int_0^1 du \frac{\sin(\pi[u+i\eta])}{\sin(\pi[u-i\eta])}e^{2i\pi(l+q)u}
\]
where $\eta =\frac{\tau}{T_0}$ is the ratio between the width of the pulse and the period of repetition.
Taking the limit $\eta\to 0$ yields
\begin{align*}
p_l 
&= \int_0^1 du \, e^{2 i \pi (l+q) u} \\
&= \frac{1}{2 i \pi (l+q)} \left( e^{2 i \pi (l+q) } - 1 \right) \\
&= \frac{\sin(\pi q)}{\pi (l+q)} \, e^{i \pi (l+q)} .
\end{align*}
Note that the case where $q$ is an integer and $l=q$ needs to be considered separately: it gives directly $p_l=1$, which is fully consistent with the general formula. 
Taking the modulus square, we obtain Eq.\,\eqref{eq8:pn_short}
\[
|p_l|^2 = \frac{\sin(\pi q)^2}{\pi^2(l+q)^2}
\]



\end{document}